\pdfoutput=1
\documentclass[11pt]{article}
\usepackage{times}
\usepackage{latexsym}
\usepackage[T1]{fontenc}
\usepackage[utf8]{inputenc}
\usepackage{microtype}
\usepackage{inconsolata}
\usepackage{bussproofs}
\usepackage{amsmath}
\usepackage{amssymb, mathrsfs}
\usepackage{tikz}
\usepackage{pgfplots}
\usepackage{subcaption}
\usepackage{tikz-dependency}
\usepackage{hyperref}
\pgfplotsset{compat=1.17}
\usetikzlibrary{positioning}

\newcommand{\pvariable}{{\bf p}}
\newcommand{\tvariable}{{\bf t}}

\newcommand{\dset}{{\bf d}}
\newcommand{\aset}{{\bf A}}
\newcommand{\cset}{{\bf C}}

\title{\bf A Categorization of Complexity Classes for Information Retrieval and Synthesis Using Natural Logic \\ 
\vspace{25pt}
}
\author{
    {\Large Greg Coppola}
    \\
    {\em coppola.ai} \\
    Research. Develop. Meme.
}
\date{\today}
\begin{document}
\maketitle
\begin{abstract}
\noindent
Given the emergent reasoning abilities of large language models, \emph{information retrieval} is becoming more complex.
Rather than just \emph{retrieve a document}, modern information retrieval systems adverstise that they can \emph{synthesize an answer} based on potentially many different documents, conflicting data sources, and using \emph{reasoning}.
But, different kinds of questions have different answers, and different answers have different complexities.
In this paper, we introduce a novel framework for analyzing the complexity of a question answer based on the \emph{natural deduction calculus} as presented in \cite{Prawitz1965}.
Our framework is novel both in that no one to our knowledge has used this logic as a basis for complexity classes, and also in that no other existing complexity classes to these have been delineated using any analagous methods either.
We identify three decidable fragments in particular called the \emph{forward}, \emph{query} and \emph{planning} fragments, and we compare this to what would be needed to do proofs for the \emph{complete} first-order calculus, for which theorem-proving is long known to be undecidable.
\end{abstract}

\tableofcontents
\section{Introduction}
\paragraph{Information Retrieval Now Requires Theorem-Proving}
Given the reasoning abilities of large language models, \emph{information retrieval} is becoming more complex.
Rather than just retrieve a document, modern information retrieval systems are asked to \emph{synthesize an answer} based on multiple data sources.
One major information company researches both how to find a ``needle in a haystack,'' as well as how to do ``geometry'' as responses to queries.

\paragraph{A Framework Based on Natural Deduction}
In this paper, we seek a \emph{general framework} for understanding the \emph{complexity classes} of different kinds of \emph{answers} that a user can ask for.
Our analysis is based on the rules for the \emph{natural deduction calculus} as presented in \cite{Prawitz1965}.
In \cite{Coppola2024}, we showed how to implement a \emph{logical graphical model}, in which each node in a \emph{Bayesian Network} is a sentence in a logical language, and each has a probability attached.
We asserted that a \emph{forward pass} in our network corresponded to inference using a \emph{specific subset} of the rules in \cite{Prawitz1965}'s calculus.
We now expand on this.

\paragraph{Analyzing the Reasoning Abilities of Transformers}
A major theme recently has become the realization that transformers are reasoning to some extent, and to investigation as to the limits of their reasoning abilities, e.g. \cite{almeekam2023planning, kambhampati2023can}.
We take the perspective that, if the end result is \emph{reasoning}, then the optimal way to study this is through formal logic, especially the \emph{first-order logic}, which is taken for granted to be sufficient to describe science and mathematics from the perspective of the philosophy of science (see, e.g. \cite{Pelletier2000}).
Our insight is that natural and useful classes and models of computation can be derived from the \emph{natural} calculus, corresponding to the difference in modern systems between doing a \emph{forward pass} through a grapical model, \emph{querying}, and \emph{planning}.

\section{Background}
\subsection{First-Order Theorems}
A {\em theorem} is a pair $\dset = \left( \aset, \cset \right)$, where $\aset$ is set of {\em assumptions} and $\cset$ is a set of {\em conclusions}, such that each $\aset$ and $\cset$ is a subset of the possible {\em sentences} in a {\em logical language} $\ell$ of interest.
A {\em proof} of $\dset$ in the calculus $\Gamma_\ell$ is a {\em sequence} of {\em deduction steps} $\gamma_\dset = [d_1, ..., d_n]$, that {\em derive} $\dset$.
Given a theorem $\dset$ and a sequence $\gamma_\dset$, we can trivially verify whether $\gamma_\dset$ {\em constitutes a valid proof} of $\dset$.
In such a case, we say that $\left(\aset, \cset\right)$ is {\em provable} in $\Gamma_\ell$.
We say that a set $\cset$ is {\em true} given $\aset$, if {\em every model satisfying} $\aset$, that honors the meaning of the {\em logical connectives} ($\land$, $\lor$, $\rightarrow$, $\forall$, $\exists$ and $\bot$), must also satisfy $\cset$.
The {\em first-order} calculus is so useful because it is {\em consistent}, meaning that everything {\em provable} is {\em true} and {\em complete}, meaning that everything {\em true} is {\em provable} \cite{Godel1931, Gentzen1934}.

\subsection{The Church-Turing Thesis}
While the task of {\em verifying} whether $\gamma_\dset$ is a valid proof of $\dset$ is trivial, the task of {\em deciding} whether $\dset$ {\em has} a proof is related to the {\em halting problem}, and is undecidable in general \cite{Turing1937, Church1936}.
That is, there is no universal program that can take an {\em arbitrary theorem} $\left( \aset, \cset \right)$ and say whether it has a proof.
If we remove the quantifiers $\forall$ and $\exists$, we are left with the {\em propositional calculus}.
Proving a theorem in this calculus corresponds to deciding {\em boolean satisfiability}, and this {\em is decidable}, but is {\em NP-hard} in general \cite{Cook1971}, which is to say $\Omega(2^N)$ where $N$ is the number of boolean variables.
The Church-Turing thesis is that anything that is \emph{computable} is computable by a machine isomorphic to a Turing machine.
Since a Turing machine's program is isomorphic to a theorem, we propose to use the theory of \emph{theorem-proving} to identify natural computing complexity classes.

\subsection{The Natural Deduction Calculus}
\cite{Prawitz1965} develops the {\em natural deduction calculus} of \cite{Gentzen1934}.
This is a \emph{complete} and \emph{consistent} calculus for first-order logic, in which there are \emph{twelve} rules, one \emph{Introduction} and one \emph{Elimination} rule for each of the six logical connectives $\land$, $\lor$, $\rightarrow$, $\forall$, $\exists$ and $\bot$.
In \cite{Coppola2024}, we show how to implement a subset of these inference rules, which we will call the \emph{Forward Fragment} in Section \ref{s:forward}.
It was always wondered why the \emph{natural deduction} calculus was considered \emph{natural}, but intuitively it did seem that there was something \emph{natural} about it.
Our work in \cite{Coppola2024} suggests a reason for this: the logic is \emph{natural} because it corresponds to how we would make a graphical model.
That is, inferences in a graphical model structure correspond to certain of the rules in \cite{Prawitz1965}'s calculus.
The rules of the natural deduction calculus are reviewed in detail in the Appendeix in Section \ref{s:natural}.
\section{The Forward Fragment}
\label{s:forward}
In this section, we investigate the \emph{Forward Fragment} of the deduction calculus.
This is the fragment implemented in \cite{Coppola2024}, and corresponds to a single forwards pass through a graphical network, in which the nodes are indexed by sentences in the first-order language.
\subsection{Fragment Definition}
\subsubsection{Horn Clauses}
Consider the first-order language with a set of predicate symbols, function symbols, and constants.
Suppose that that we have $n$ different functions $P_i$ of at most $k$ variables, and suppose that $C$ is a function of $k$ open variables.
Then, the \emph{Forward} fragment, implemented in \cite{Coppola2024}, involves a system of \emph{quantified Horn clauses} of the form:
\begin{equation}
\forall x_1, \ldots, \forall x_k, \Bigg[ \bigwedge_{i=1}^{n} P_i(x_1, \ldots, x_k) \Bigg] \rightarrow C(x_1, \ldots, x_k)
\label{e:basic_horn}
\end{equation}
We emphasize that every variable mentioned $x_1, ..., x_k$ must appear in $C$.
This is called the \emph{safety} restriction of \emph{Datalog}, which is very crucial to the distinction in complexity classes, and is further discussed in Section \ref{s:datalog}.
An example of an inference that uses the form is the following:
\begin{equation}
    \forall x_1, x_2, \Bigg[like(x_1, x_2) \wedge like(x_2, x1) \Bigg] \rightarrow friends(x_1, x_2)
\end{equation}
This says that if any $x_1$ and $x_2$ both like \emph{each other}, then they are \emph{friends}.

\subsubsection{Disjunctive Normal Form}
The presence of multiple ways to derive the same predicate amounts to a \emph{disjunction} over the possible ways to reach that conclusion, and thus $m$ statements of the form \ref{e:basic_horn} can be written as a statement with a disjoined premise as:
\begin{equation}
\forall x_1, \ldots, \forall x_k, \Bigg[ \bigvee_{i=1}^{m} \bigwedge_{i=1}^{n} P_{i,j}(x_1, \ldots, x_k) \Bigg] \rightarrow C(x_1, \ldots, x_k)
\label{e:dnf_form}
\end{equation}
Thus, any sentence in \emph{disjunctive normal form}, apart from the restriction on quantification, can be used as a premise.
Since any statement in the propositional calculus can be written in disjunctive normal form \cite{andrews1986introduction}, any statement obeying the restriction on quantification can be a premise to a deduction rule, so this is a very powerful framework.

\subsubsection{Conjoined Conclusions}
Finally, if we have a statements of the form \ref{e:dnf_form}, note that in some cases, the premise can be the same, and so a database of different statements like \ref{e:dnf_form} can be re-written as a single statement like:
\begin{equation}
\forall x_1, \ldots, \forall x_k, \Bigg[ \bigvee_{i=1}^{n}  \bigwedge_{i=1}^{n} P_i(x_1, \ldots, x_k) \Bigg] \rightarrow \Bigg[ \bigwedge_{i=1}^{n}  C(x_1, \ldots, x_k) \Bigg]
\label{e:final_horn}
\end{equation}
That is, a set of statements of the form \ref{e:basic_horn} and \ref{e:final_horn} are equivalent in the proofs that they allow.

\subsection{Datalog's Safety Restriction}
\label{s:datalog}
In this \emph{Forward} fragment, we are incorporating an important restriction on the forms of clauses, which is that we can only mention in the $P_i(x_1, ..., x_k)$ variables that have also been mentioned in $C(x_1, ..., x_k)$.
That is, in this fragment, we cannot introduce new universally quantified variables used in $P_i$ but not in $C$, 
and means we cannot say that we are \emph{truly} implementing \emph{$\forall$-Elimination}, for example, because the \emph{$\forall$-Elimination} rule in \cite{Prawitz1965}'s system does \emph{not} include this restriction.
A Datalog rule is considered to be \textit{safe} if every variable that appears in the head of the rule (the conclusion part) also appears in a positive literal in the body of the rule (the condition part) \cite{abiteboul1995foundations}.

\subsection{Analysis of the Fragment}
This fragment allows all statements that use only the following rules:
\emph{$\land$-Elimination},
\emph{$\land$-Introduction},
\emph{$\lor$-Introduction},
\emph{$\forall$-Elimination} and
\emph{$\rightarrow$-Elimination}, but subject to the \emph{safety} restriction on quantification as discussed above.
Because this fragment follows the safety restriction of Datalog, it amounts to \emph{Horn Satisfiability}, which is a well-known efficient fragment, where the time taken to do inference is \emph{linear} in the number of variables \emph{total} in the theory, which is really very efficient, when compared to the other fragments, as we will see.
\cite{Coppola2024} implements this fragment, and shows how we can not only do logical inference, but even assign probabilities if this fragment is stuck to.
That is, we give probabilities to a similar fragment as that for which \cite{pereira1987prolog} investigated theorem-provability.
That is, we can distinguish between theories that are more or less \emph{likely}.
And, we can determine a \emph{generative} probability for the data set, so that it can be compressed \cite{SutskeverObservation}.

\section{The Query Fragment}
\label{s:query}
\subsection{Motivation}
Suppose that instead of \ref{e:basic_horn}, suppose we have an integer $K > k$, and:
\begin{equation}
\forall x_1, \ldots, \forall x_K, \Bigg[ \bigwedge_{i=1}^{n} P_i(x_1, \ldots, x_k, x_{k+1}, ..., x_K) \Bigg] \rightarrow C(x_1, \ldots, x_k)
\label{e:unbound_universal}
\end{equation}
This statement is \emph{unsafe} in the terms of Datalog (see Section \ref{s:datalog}), because the variables $x_{k+1}, ..., x_K$ are \emph{not mentioned} in the conclusion, and thus \emph{unbound}.
Such a statement can be useful, for example, in saying that if $x_1$ and $x_2$ both want the same thing $x_3$, then $x_1$ and $x_2$ are \emph{in competition}:
\begin{equation}
\forall x_1, x_2, x_3, \Bigg[ want(x_1, x_3) \wedge want(x_2, x_3) \Bigg] \rightarrow compete(x_1, x_2)
\end{equation}
Thus, the restriction that we have put in Section \ref{s:datalog} is something that we want to relax.

\subsection{Existential Quantifiers as Queries}
We can re-write \ref{e:unbound_universal} as:
\begin{equation}
\forall x_1, \ldots, \forall x_k, \Bigg[ \exists x_{k+1}, ..., x_K, \bigwedge_{i=1}^{n} P_i(x_1, \ldots, x_k, x_{k+1}, ..., x_K) \Bigg] \rightarrow C(x_1, \ldots, x_k)
\end{equation}
Each existentially quantified variable $\exists x_{k+i}$ corresponds to a \emph{query} as to whether any element in the domain can be found that satisfies the premise.

\subsection{Complexity of the Full Fragment}
If our goal is to do \emph{full deterministic} theorem-proving, the worst-case bound is very bad, for reasons that we will now investigate.
\paragraph{Quantifiers in Same Sentence}
Suppose that the $x_i$ are universally quantified and the $y_j$ are existentially quantified, but we will still write $\exists$ for clarity.
Now, consider an implication with a single existential:
\begin{equation}
\left[ \exists y_1, P(x_1, y_1) \right] \rightarrow C(x_1)
\end{equation}
If the domain that $\exists y_1$ ranges over is finite, which we assume that it is, then we can \emph{implement} $\exists$ using \emph{$\vee$-Introduction}, with a disjunction over $D$ elements in the domain:
\begin{equation}
\left[ \bigvee_{i=1}^D (x_1, c_i) \right] \rightarrow \exists y_1 P(x_1, y_1)
\end{equation}
This is computable, but quickly becomes a problem from an efficiency perspective.
For example, to implement a premise with two exisentially quantified variables in the premise like:
\begin{equation}
\left[ \exists y_1, y_2, P(x_1, y_1, y_2) \right] \rightarrow C(x_1)
\end{equation}
This would require $\Theta(D^2)$ simple rules to implement.
In general, if we begin with a graph of size $G$, and a domain of size $D$, then to implement $N$ $\exists$ would require work $\Theta(GD^N)$.

\subsection{Useful Best-Effort Fragments}
We must stress that the exponential blow-up in computational cost described above is only in the \emph{worst-case} assuming \emph{full theorem-proving}.
We discuss two efficient but useful fragments of the query calculus that we believe can be helpful in practice.

\paragraph{Shallow Queries}
Suppose instead, we can say that, when eliminating a $\exists$, we are interested only in facts that are \emph{already in} an existing \emph{statically} stored database, without requiring any \emph{dynamic} theorem-proving.
In other words, we can make a single database query to see if there is any $x_3$ such that $want(x_1, x_3)$ and $want(x_2, x_3)$.
This can be done in $O(1)$ time relative to the complexity of the graph, possibly even using a database index.

\paragraph{Probabilistic Ranking}
In traditional, non-probabilistic theorem-proving, if we want to prove $\exists x P(x)$, we have no a priori way to \emph{order} the $x$ in terms of which would be \emph{most likely} make $P(x)$ true.
Such a ranking could cut down on the search time.
We could also use a heuristic like \emph{try only the most probable candidate}, which would eliminate the exponential blow up in the search space, by only allowing $1$ try per $\exists$.

\paragraph{A* Search}
Otherwise, if we have a \emph{ranking} of different candidates, and we really want to search exhaustively, or as exhaustively as possible, we speculate we can view this as analogous to searching through a \emph{maze} and use \emph{A* search} as in \cite{lehnert2024a}.
\section{The Planning Fragment}
\label{s:planning}
The natural rule of \emph{$\lor$-Elimination} we also call \emph{reasoning by cases}, and now we will investigate why that is, and see the relation to \emph{planning} under \emph{uncertain conditions}.
\subsection{Example of Reasoning Under Uncertainty}
\paragraph{Set-Up}
Let us consider the example of a party that is planning an \emph{enjoyable excursion}, where the goal at the end is to be \emph{satisfied} with their experience.
This particular party $\pvariable$ has their own individual preferences.
And, they are considering a trip to a particular beach town $\tvariable$.
Suppose the party will be \emph{satisfied} with their excursion if the excitement level $l$ with excursion $e$ is at least a $7$ out of $10$:
\begin{equation}
    \forall e, \exists l, \Bigg[ excited(\pvariable, e, l) \wedge l \geq 7 \Bigg] \rightarrow satisfied(\pvariable, e)
\end{equation}
Here, we assume that we can use $\exists$.
If the party visits the beach town $\tvariable$ they can go to the beach:
\begin{equation}
    \Bigg[ visit(\pvariable, e, \tvariable) \Bigg] \rightarrow visit(\pvariable, e, beach(\tvariable))
\end{equation}
If they visit the beach, and it is sunny, the happiness level will be a $10$.
\begin{equation}
    \Bigg[ visit(\pvariable, e, beach(\tvariable)) \wedge sunny(e) \Bigg] \rightarrow excited(\pvariable, e, 10)
\end{equation}
But, if at the beach, it is not sunny, the party will be \emph{very} unhappy:
\begin{equation}
    \Bigg[ visit(\pvariable, e, beach(\tvariable)) \wedge \neg sunny(e) \Bigg] \rightarrow excited(\pvariable, e, 1)
\end{equation}
Now, there is a favorite \emph{restaurant} of the party's in \tvariable, which they enjoy going to when it rains:
\begin{equation}
    \Bigg[ visit(\pvariable, e, restaurant(\tvariable)) \wedge \neg sunny(e) \Bigg] \rightarrow excited(\pvariable, e, 7)
\end{equation}
Going to this restaurant is not as fun as the beach, but it is still fun on a rainy day.
However, going to the restaurant if it is \emph{not} sunny does not make the party happy, as they will wonder why they are not at the beach:
\begin{equation}
    \Bigg[ visit(\pvariable, e, restaurant(\tvariable)) \wedge sunny(e) \Bigg] \rightarrow excited(\pvariable, e, 3)
\end{equation}
\paragraph{Observation}
We are now in a situation where the party actually \emph{can} guarantee that they will be satisfied with their excursion, becuase they can go to the town $\tvariable$, and if it is sunny, they can go to the beach, and if it rains, they can go to their restaurant, and they will be happy either way.
The fact that they will be happy \emph{either way} requires the ability to reason by \emph{cases}, and evaluate their position \emph{in either case}.

\subsection{Disjunctive Normal Form}
In order to handle the motivating case, and others as well, it suffices to allow, instead of \ref{e:dnf_form}, that a conclusion can now itself contain a disjunction as in:
\begin{equation}
\forall x_1, \ldots, \forall x_k, \Bigg[ \bigvee_{i=1}^{n} \bigwedge_{i=1}^{n} P_i(x_1, \ldots, x_k) \Bigg] \rightarrow \bigvee_{i=1}^{n} C(x_1, \ldots, x_k)
\label{s:planning_form}
\end{equation}
Of course, it is less information to learn that $A \lor B$ than to learn $A$, because the latter allows us to conclude $A$, and the former does not, and this is why we must \emph{resaon by cases}.

\subsection{Two-Player Games}
In general, interaction between the agent and the environment can be viewed as a two-player game, where the agent plays themselves, and \emph{reality} plays as the opponent.
In an \emph{antagoistic} two-player game, we assume that the \emph{opponent} always does their best to make the life of the \emph{protagonist} as bad as possible.
In the case of the \emph{protagonist} against \emph{reality}, \emph{reality} might not always try to make things as bad as possible for the \emph{protagonist}, but rather reality will react according to a \emph{probability distribution}.
In either case, this is similar to a game of \emph{chess} or \emph{go} \cite{silver2016mastering}.

\subsection{Relevant Empirical Results}
Because of the $\Omega(2^N)$ explosion in complexity of solving a boolean satisfiability problem \cite{Cook1971}, we expect that a \emph{large language model} of finite size will not be able to handle problems corresponding to boolean satisfiability, if the input grows large enough.
This has been investigated and shown in \cite{almeekam2023planning, kambhampati2023can}.

\section{Discussion}
\subsection{Review of Studied Fragments}
In the \emph{forward} fragment of Section \ref{s:forward}, we investigate a system that contains the rules:
\begin{itemize}
    \item {\bf forward fragment}
    \begin{itemize}
        \item \emph{$\land$-Introduction}
        \item \emph{$\land$-Elimination}
        \item \emph{$\lor$-Introduction}
        \item \emph{$\rightarrow$-Elimination}
        \item \emph{$\forall$-Elimination} (limited by the \emph{safety} requirement on quantification)
    \end{itemize}
\end{itemize}
In the \emph{query} fragment of Section \ref{s:query}, we weaken the restriction on quantification and add \emph{$\exists$-Introduction}:
\begin{itemize}
    \item {\bf query fragment}
    \begin{itemize}
        \item all those from the \emph{direct fragment}, plus
        \item \emph{$\forall$-Elimination} (without limitation)
        \item \emph{$\exists$-Introduction}
    \end{itemize}
\end{itemize}
Adding in \emph{reasoning by cases} creates the \emph{planning fragment} of Section \ref{s:planning}:
\begin{itemize}
    \item {\bf planning fragment}
    \begin{itemize}
        \item all those from the \emph{query fragment}, plus
        \item \emph{$\vee$-Elimination}
    \end{itemize}
\end{itemize}

\subsection{The Remaining Fragments}
The remaining fragments are the ones that \cite{Prawitz1965} called \emph{improper}:
\begin{itemize}
    \item {\bf improper rules}
    \begin{itemize}
        \item \emph{$\lor$-Elimination}
        \item \emph{$\rightarrow$-Introduction}
        \item \emph{$\forall$-Introduction}
        \item \emph{$\exists$-Elimination}
        \item \emph{$\bot$-Introduction}
    \end{itemize}
\end{itemize}
In \cite{Coppola2024} we implemented the \emph{direct} fragment, and showed how to assign probabilities to conclusions.
We said that one feature of the research is that it shows how to \emph{unify logical and probabilistic reasoning}, by providing a \emph{unified model} of the two.
This is because, while we can show what happens in a \emph{forward pass} of a \emph{direct} fragment, we can also show how this fragment that is implemented related to a \emph{complete} and \emph{consistent} calculus for the first-order logic.
However, it is important to understand the reason that we can \emph{not} implement an entire logic as a \emph{forward pass} in a graphical network.
To do proofs in the unbound fragments, one must implement \emph{partial} strategies for these rules that \cite{Prawitz1965} called \emph{improper}.
Such a list must always be incomplete, because full theorem proving in the first-order calculus is undecidable in general \cite{Church1936, Turing1937}.
\section{Appendix: The Natural Deduction Calculus}
\label{s:natural}
\paragraph{Introduction}
This is a review of \cite{Prawitz1965}'s formulation of the \emph{natural deduction calculus}.
\cite{Prawitz1965} distinguishes between deduction rules that are \emph{proper} and \emph{improper}.
We instead call them \emph{simple} and \emph{complex}, to avoid the value judgment.
We will point out for each rule which category \cite{Prawitz1965} gave it.
The \emph{complex} ones are harder to implement.

\paragraph{The Logical Rules}
This is a \emph{complete} and \emph{consistent} calculus for first-order logic, in which there are \emph{twelve} rules, one \emph{Introduction} and one \emph{Elimination} rule for each of the six logical connectives $\land$, $\lor$, $\rightarrow$, $\forall$, $\exists$ and $\bot$.
\paragraph{$\land$-Introduction}
The rule of {\em $\land$-Introduction} says that if we have proved $A$ and $B$, we can conclude $A \land B$:
\begin{equation}
    \left\{\left(\aset_0, \cset_0 \cup \left\{A, B\right\}\right)\right\} \rightarrow \left(\aset_0, \cset_0 \cup \left\{A \wedge B\right\}\right)
\end{equation}
This inference rule is \emph{simple}.

\paragraph{$\land$-Elimination}
The rule of {\em $\land$-Elimination} says that if we have proved $A \land B$, we can conclude both $A$ and $B$:
\begin{equation}
    \left\{\left(\aset_0, \cset_0 \cup \left\{A \wedge B\right\}\right)\right\} \rightarrow \left(\aset_0, \cset_0 \cup \left\{A, B\right\}\right)
\end{equation}
This inference rule is {\em simple}.
Note that $\land$ is the only logical symbol for which {\em both} rules are simple.

\paragraph{$\lor$-Introduction}
The rule of {\em $\lor$-Introduction} says that if we have either $A$ or $B$, we can conclude $A \lor B$:
\begin{equation}
    \left\{\left(\aset_0, \cset_0 \cup \left\{A\right\}\right)\right\} \rightarrow \left(\aset_0, \cset_0  \cup \left\{A\right\} \cup \left\{A \lor B\right\}\right)
\end{equation}
This inference rule is {\em simple}.
This corresponds to a {\em disjunction} gate in the {\em QBBN}, and can be {\em learned} to make a statistical model \cite{Coppola2024}.

\paragraph{$\lor$-Elimination}
{\em $\lor$-Elimination} says that if we have proved $A \lor B$, and we have proven that from $A$ we can conclude $C$, and we have proven that from $A$ we can conclude $C$, then we can conclude $C$.
\begin{equation}
    \left\{
    \begin{aligned}
        &\left(\aset_0 \cup \left\{\right\}, \cset_0 \cup \left\{A \lor B\right\}\right) \\
        &\left(\aset_0 \cup \left\{A\right\}, \cset_0 \cup \left\{C\right\}\right) \\
        &\left(\aset_0 \cup \left\{B\right\}, \cset_0 \cup \left\{C\right\}\right) 
    \end{aligned}
    \right\}
    \rightarrow \left(\aset_0, \cset_0 \cup \left\{C\right\}\right)
\end{equation}
This is {\em complex}, and actually involves {\em three} sets of assumptions, $\aset_0$, $\aset_0 \cup \left\{A\right\}$ and $\aset_0 \cup \left\{B\right\}$.
Although this deduction is complex, we can show in a sense that this rule is less complex than the others, in that we can always decide proofs in a logic that only uses the simple rules and this rule, but it becomes NP-hard, see Section \ref{s:planning}.

\paragraph{$\rightarrow$-Introduction}
The rule of {\em $\rightarrow$-Introduction} says that if we know we can prove $B$ from $A$, then we can prove $A\rightarrow B$:
{\em $\rightarrow$-Introduction} is {\em complex} and says:
\begin{equation}
    \left\{\left(\aset_0 \cup \left\{A\right\}, \cset_0 \cup \left\{B\right\}\right)\right\} \rightarrow \left(\aset_0 \cup \left\{\right\}, \cset_0 \cup \left\{A \rightarrow B\right\}\right)
\end{equation}
This is {\em complex} because it involves a change of assumptions from $\aset_0 \cup \left\{A\right\}$ to $\aset_0$.
This is a rare rule to use in practice.

\paragraph{$\rightarrow$-Elimination}
The rule of {\em $\rightarrow$-Elimination} is {\em simple}. It corresponds to a single factor forward inference in the {\em QBBN}.
Starting with the {\em simple} {\em $\rightarrow$-Elimination}, we have:
\begin{equation}
    \left\{\left(\aset_0, \cset_0 \cup \left\{A, A \rightarrow B\right\}\right)\right\} \rightarrow \left(\aset_0, \cset_0  \cup \left\{A, A \rightarrow B\right\} \cup \left\{B\right\}\right)
\end{equation}
This is simple.
This deduction is the bedrock of logic, i.e., {\em modus ponens}.

\paragraph{$\forall$-Introduction}
The rule of {\em $\forall$-Introduction} asserts that if we can prove \(A\) without specifically referencing any particular instance of \(x\), then \(A\) can be generalized to \(\forall x A\):
\begin{equation}
    \left\{\left(\aset_0, \cset_0 \cup \{A\}\right)\right\} \rightarrow \left(\aset_0, \cset_0 \cup \left\{\forall x.A\right\}\right)
\end{equation}
This rule is considered {\em complex} due to the introduction of a universal quantifier, expanding the scope of \(A\) to all instances of \(x\). The assumptions remain unchanged.

\paragraph{$\forall$-Elimination}
The rule of {\em $\forall$-Elimination} allows for the instantiation of a universally quantified statement to a specific instance. Given \(\forall x A\), we can deduce \(A[t/x]\) for any term \(t\):
\begin{equation}
    \left\{\left(\aset_0, \cset_0 \cup \left\{\forall x.A\right\}\right)\right\} \rightarrow \left(\aset_0, \cset_0 \cup \left\{A[t/x]\right\}\right)
\end{equation}
This rule is {\em simple}.

\paragraph{$\exists$-Introduction}
The rule of {\em $\exists$-Introduction} posits that if \(A[t/x]\) is proven for some term \(t\), then \(\exists x A\) is also proven:
\begin{equation}
    \left\{\left(\aset_0, \cset_0 \cup \left\{A[t/x]\right\}\right)\right\} \rightarrow \left(\aset_0, \cset_0 \cup \left\{\exists x.A\right\}\right)
\end{equation}
This rule is {\em simple}.

\paragraph{$\exists$-Elimination}
The rule of {\em $\exists$-Elimination} involves deducing a conclusion \(B\) from an existential premise \(\exists x.A\), requiring a temporary assumption \(A[c/x]\) for a new constant \(c\):
\begin{equation}
    \left\{\left(\aset_0 \cup \left\{\exists x.A\right\}, \cset_0\right), \left(\aset_0 \cup \left\{A[c/x]\right\}, \cset_0\right) \right\} \rightarrow \left(\aset_0, \cset_0 \cup \left\{B\right\}\right)
\end{equation}
This rule is {\em complex} because it involves the introduction of a new assumption for a hypothetical instance \(c\) that satisfies \(A\), and the set of assumptions changes during the process.

\paragraph{$\bot$-Introduction}
The rule of {\em $\bot$-Introduction} allows for the introduction of a contradiction (\(\bot\)) when both a statement \(A\) and its negation \(\neg A\) are proven:
\begin{equation}
    \left\{\left(\aset_0, \cset_0 \cup \left\{A, \neg A\right\}\right)\right\} \rightarrow \left(\aset_0, \cset_0 \cup \left\{\bot\right\}\right)
\end{equation}
This rule is {\em complex} as it derives a contradiction, indicating a fundamental inconsistency within the assumptions.

\paragraph{$\bot$-Elimination}
Applying the rule of {\em $\bot$-Elimination}, any statement \(B\) can be concluded from a contradiction (\(\bot\)):
\begin{equation}
    \left\{\left(\aset_0, \cset_0 \cup \left\{\bot\right\}\right)\right\} \rightarrow \left(\aset_0, \cset_0 \cup \left\{B\right\}\right)
\end{equation}
This rule is {\em simple}.

\bibliographystyle{apalike}
\bibliography{bibtex}
\end{document}